\RequirePackage{fix-cm}
\documentclass[smallextended]{svjour3}       
\smartqed  
\usepackage{graphicx,epsf,epsfig,amssymb} 
\usepackage{amsmath,amssymb}
\usepackage{amsfonts}
\usepackage{xspace} 
\usepackage[usenames]{color}
\usepackage{dcolumn}
\usepackage{bm}
\usepackage{mathrsfs}
\usepackage{aas_macros} 
\usepackage[colorlinks=true]{hyperref}
\usepackage[all]{hypcap} 
\usepackage[utf8]{inputenc} 
\usepackage{cite} 
\newcommand{\eq}{\begin{equation}}
\newcommand{\be}{\begin{equation}}
\newcommand{\eeq}{\end{equation}}
\newcommand{\ee}{\end{equation}}
\newcommand\ba{\begin{eqnarray}}
\newcommand\ea{\end{eqnarray}}

\begin{document}

\title{Topical Collection: Testing the Kerr spacetime with gravitational-wave and electromagnetic observations}

\titlerunning{}        

\author{Emanuele Berti}


\institute{E. Berti \at
Department of Physics and Astronomy, Johns Hopkins University, 3400 N. Charles Street, Baltimore, Maryland, 21218, USA}

\date{Received: date / Accepted: date}

\maketitle

\begin{abstract}
One of the Holy Grails of observational astronomy is to confirm the prediction that black holes in the Universe are described by the Kerr solution of Einstein's field equations of general relativity. This Topical Collection provides a status report of theoretical and experimental progress towards confirming the ``Kerr paradigm'' through X-ray astronomy, gravitational lensing, stellar tidal disruption events, superradiance, and gravitational-wave observations of black hole binary mergers.
\end{abstract}


\vspace{.5cm}


The LIGO/Virgo gravitational wave detections, the imaging of a black hole shadow by long-baseline radio interferometry, and ever more precise observations across all electromagnetic wavelengths have opened an unprecedented window onto gravity at its strongest. Just as a wealth of experimental data has revolutionized cosmology by confronting theoretical speculations with measurement, strong-field gravity is expanding beyond the domain of mathematical physics to become a precision experimental science.

Black holes -- the simplest, most compact, and physically elusive macroscopic objects in the Universe -- play a central role in this new era in physics and astronomy.  Among astronomical targets, black holes are extraordinary in their ability to convert energy into electromagnetic and gravitational radiation. From a theoretical point of view, the strong-field dynamics of black holes challenges our understanding of nonlinear partial differential equations, numerical methods, and quantum field theory in curved spacetime.  The main conceptual problems in black hole physics hold the key to fundamental issues in theoretical physics: for example, the black hole information paradox and the Hawking-Penrose singularity theorems point to deep inconsistencies in our current understanding of gravity and quantum mechanics.

This is why the observational confirmation of the ``Kerr paradigm'' -- the expectation that black holes in the Universe are described by the vacuum solution of the field equations of general relativity found by Roy Kerr~\cite{Kerr:1963ud} -- has become a Holy Grail of modern observational astronomy. This Topical Collection presents articles reviewing theoretical and observational developments in this direction.

Krawczynski~\cite{Krawczynski:2018fnw} reviews progress in X-ray observations of stellar-mass and supermassive black holes. While there is already clear evidence for relativistic effects in strong-field gravity, quantitative tests of the Kerr hypothesis are still limited by theoretical and practical difficulties: astrophysical uncertainties are still large compared to the observational differences between the Kerr metric and proposed alternatives. Further progress will be possible through numerical simulations and improvements in X-ray spectroscopy, timing, polarimetry, and interferometry.

Light rings and fundamental photon orbits play a central role in strong gravitational lensing around compact objects and in the description of black hole shadows. Cunha and Herdeiro~\cite{Cunha:2018acu}
overview the theoretical foundations and phenomenology of fundamental photon orbits in general relativity and in modified theories of gravity, stressing how these concepts can be used to test the Kerr hypothesis.

These theoretical foundations underlie the interpretation of present and future data by the Event Horizon Telescope, a millimeter VLBI array that aims to take horizon-scale resolution ``pictures'' of the black holes in the center of the Milky Way (Sgr~A$^*$) and of the M87 galaxy. Psaltis~\cite{Psaltis:2018xkc} shows that measurements of the shape and size of the shadows could lead to black hole spin measurements, test the cosmic censorship conjecture and no-hair theorems, and even provide evidence for classical effects of the quantum structure of black holes. For Sgr~A$^*$, measurements of the precession of stellar orbits and the timing of orbiting pulsars offer complementary avenues to the gravitational tests with the Event Horizon Telescope.

Stone et al.~\cite{Stone:2018nbx} consider yet another interesting probe of the Kerr metric: tidal disruption events (TDEs), which occur when a star passes too close to a supermassive black hole and is ripped apart by its tidal field, powering a bright electromagnetic flare as it is accreted by the black hole. Several dozen TDE candidates have been observed.  When the black hole mass is larger than $\sim 10^7M_\odot$, TDEs occur within ten gravitational radii of the hole. This gives us observational access to general relativistic effects, including: (1) a super-exponential cutoff in the volumetric TDE rate for black hole masses above $\sim 10^8M_\odot$ due to direct capture of tidal debris by the event horizon; (2) delays in accretion disk formation (and a consequent alteration of the early-time light curve) caused by the effects of relativistic nodal precession on stream circularization; and (3) quasi-periodic modulation of X-ray emission due to global precession of misaligned accretion disks and the jets they launch.

One of the definining properties of the Kerr metric is the existence of an ``ergoregion'': a region of spacetime located outside of the event horizon where no time-like observer could remain stationary. In 1969, Roger Penrose showed that particles within the ergosphere can have negative energy as measured by an observer at infinity. When captured by the horizon, these negative-energy particles extract mass and angular momentum from the black hole. The decay of a single particle within the ergosphere is not a particularly efficient means of energy extraction, but -- as reviewed by Schnittman~\cite{Schnittman:2018ccg}
-- the collision of multiple particles can reach arbitrarily high center-of-mass energy in the limit of extremal black hole spin. The resulting particles can escape with high efficiency, potentially serving as a probe of high-energy particle physics and of general relativity. The collisional Penrose process could enhance the annihilation of dark matter particles in the vicinity of a supermassive black hole, and therefore allow us to shed light on the nature of dark matter.

Last but not least, gravitational-wave interferometers such as LIGO and Virgo now allow us to test strong-field general relativity and the Kerr paradigm in completely new ways. Two reviews in this collection~\cite{Berti:2018cxi,Berti:2018vdi} discuss present and future gravitational-wave bounds on deviations from general relativity and tests of the dynamics of Kerr black holes. These tests involve either theory-agnostic frameworks (such as the parametrized post-Einsteinian formalism for inspiral/merger radiation, and parametrized ringdown approaches to describe the post-merger waveforms) or specific modified theories of gravity (such as scalar–tensor, Lorentz-violating and extra dimensional theories, Einstein–scalar–Gauss–Bonnet and dynamical Chern–Simons gravity).

\begin{acknowledgements}

E.B. is supported by NSF Grant No. PHY-1912550, NSF Grant No. AST-1841358, NSF-XSEDE Grant No. PHY-090003, and NASA ATP Grant No. 17-ATP17-0225. 
The author would like to acknowledge networking support by the GWverse COST Action CA16104, ``Black holes, gravitational waves and fundamental physics.''
\end{acknowledgements}

\bibliographystyle{spphys}       
\bibliography{master}   

\end{document}